# Investigating the genomic background of CRISPR-Cas genomes for CRISPR-based antimicrobials

Hyunjin Shim[1,†]

*[1]Center for Biosystems and Biotech Data Science, Ghent University Global Campus, Songdo, ICN, Republic of Korea*

[†]Corresponding Author: Hyunjin Shim (jinenstar@gmail.com)

## Abstract

CRISPR-Cas systems are an adaptive immunity that protects prokaryotes against foreign genetic elements. Genetic templates acquired during past infection events enable DNA-interacting enzymes to recognize foreign DNA for destruction. Due to the programmability and specificity of these genetic templates, CRISPR-Cas systems are potential alternative antibiotics that can be engineered to self-target antimicrobial resistance genes on the chromosome or plasmid. However, several fundamental questions remain to repurpose these tools against drug-resistant bacteria. For endogenous CRISPR-Cas self-targeting, antimicrobial resistance genes and functional CRISPR-Cas systems have to co-occur in the target cell. Furthermore, these tools have to outplay DNA repair pathways that respond to the nuclease activities of Cas proteins, even for exogenous CRISPR-Cas delivery. Here, we conduct a comprehensive survey of CRISPR-Cas genomes. First, we address the co-occurrence of CRISPR-Cas systems and antimicrobial resistance genes in the CRISPR-Cas genomes. We show that the average number of these genes varies greatly by the CRISPR-Cas type, and some CRISPR-Cas types (IE and IIIA) have over 20 genes per genome. Next, we investigate the DNA repair pathways of these CRISPR-Cas genomes, revealing that the diversity and frequency of these pathways differ by the CRISPR-Cas type. The interplay between CRISPR-Cas systems and DNA repair pathways is essential for the acquisition of new spacers in CRISPR arrays. We conduct simulation studies to demonstrate that the efficiency of these DNA repair pathways may be inferred from the time-series patterns in the RNA structure of CRISPR repeats. This bioinformatic survey of CRISPR-Cas genomes elucidates the necessity to consider multifaceted interactions between different genes and systems, to design effective CRISPR-based antimicrobials that can specifically target drug-resistant bacteria in natural microbial communities.

## Introduction

Clustered regularly interspaced short palindromic repeats (CRISPR), found in many prokaryotic genomes, store sequence information about foreign DNA that has invaded these microorganisms [1–3]. With this information, the CRISPR-associated system (Cas genes) provide an adaptive immunity that protects the cell against invasive mobile genetic elements such as bacteriophages. The ability of CRISPR-Cas systems to cut and edit DNA has opened a new era of genome-editing technologies in various fields such as medicine and agriculture [4]. Such applications have driven the scientific community to discover diverse CRISPR-Cas systems in nature, to uncover those that may be better tools for editing eukaryotic genomes [5–7]. CRISPR-Cas systems are currently divided into Class 1 (Type I, III, IV) and Class 2 (Type II, V, VI), with each type further classified into several subtypes [5].

CRISPR-Cas systems are recently being investigated for their potential to selectively target bacteria with antimicrobial resistance (AMR) genes [8–10]. Antimicrobial resistance is now considered a "hidden pandemic" which threatens to undermine the effectiveness of modern medicine, from minor surgical procedures to cancer treatments due to hospital-acquired infections [11]. In 2019, infections from multidrug-resistant bacteria were estimated to have caused more than 1.2 million deaths worldwide [12]. Given the severity of the uncontrolled spread of these superbugs, the World Health Organization (WHO) recently published a list of priority pathogens which urgently need new antibiotics, including carbapenem-resistant *Acinetobacter baumannii* and *Pseudomonas aeruginosa*. CRISPR-based antimicrobials are potential alternatives to the traditional small-molecule antibiotics, as the CRISPR component is programmable to target specific genes with a complex of Cas proteins. Several studies independently engineered CRISPR-Cas systems to selectively remove AMR genes from bacterial populations [13–15].

Despite the potential of CRISPR-based antimicrobials, several challenges remain before these tools can be successfully repurposed to remove AMR-carrying bacteria or plasmids from natural microbial communities [9,10,16]. In addition to the practical issues such as the delivery to target bacteria, there are several fundamental questions related to the effectiveness of CRISPR-based antimicrobials. For endogenous CRISPR-Cas self-targeting, both AMR genes and functional CRISPR-Cas systems have to be present in the chromosome or plasmid of target bacteria. For such bacteria, CRISPR-based antimicrobials can simply be composed of a self-targeting CRISPR array that is compatible with the endogenous Cas system [13,17]. Without functional endogenous CRISPR-Cas systems, a complete set of CRISPR-Cas systems that targets a specific AMR gene has to be delivered exogenously [8]. Thus, it is necessary to understand the genomic background of target bacteria for effective design and delivery of CRISPR-based antimicrobials. In this study, we use the public CRISPR-Cas database to survey the genomic background of CRISPR-Cas genomes, which we define as prokaryotic genomes that have one functional CRISPR-Cas system (Figure 1). These CRISPR-Cas genomes are searched for AMR genes, to investigate the co-occurrence of functional CRISPR-Cas systems and AMR genes in diverse bacteria, particularly in pathogenic bacteria.

Another pertinent question is the impact of DNA repair pathways on the effectiveness of CRISPR-based antimicrobials [8–10]. Bacteria have evolved complex DNA repair pathways that can repair DNA damage in response to various external and internal triggers (e.g. UV irradiation, antibiotics, stalled replication, recombination) that can be lethal if not repaired before cell division [18,19]. Despite the high efficiency of self-targeting spacers, a small percentage of the bacterial population targeted by these CRISPR-Cas systems persisted in a number of previous studies [13,20]. Here, we scan the CRISPR-Cas genomes for DNA repair pathways to investigate the potential interference against the activities of CRISPR-based antimicrobials.

We further explore the interplay of CRISPR-Cas systems and DNA repair pathways through simulation studies, whose co-evolution was predicted by the Lamarckian evolution of directed mutagenesis [21,22]. It is intriguing to observe that the acquisition of new spacers in CRISPR arrays requires DNA repair, during which several proteins engage in DNA unwinding, editing, and repairing activities along with the Cas proteins. Recent evidence shows that most CRISPR-Cas systems acquire new spacers through site-specific integration, with the leader end spacers being the most recent and most active [23–25]. This strategy enables prioritizing the defense against the most recent invader at the leader end by differential expression of crRNAs across the CRISPR array. However, this acquisition step is susceptible to mutation accumulation in the CRISPR repeats without efficient DNA repair pathways. Thus, we investigate the time-series patterns in CRISPR repeats to examine the potential interference of DNA damage response in utilizing CRISPR-based antimicrobials against prokaryotes. We first examine how the RNA structures of CRISPR repeats change over time by visualizing and analyzing the time-series patterns of CRISPR arrays associated with different Cas system types. We show that Class 1 CRISPR repeats are more structured than Class 2 CRISPR repeats, and this structural component is maintained throughout the site-specific integration of new spacers over time, indicating the active role of DNA repair pathways in these genomes. Furthermore, we show that DNA repair pathways in these CRISPR-Cas genomes are numerous and diverse. These results demonstrate that the genomic background of target bacteria should be considered for DNA damage response for effective design of CRISPR-based antimicrobials tailored against these disease-causing strains.

## Results

### CRISPR-Cas genomes have numerous antimicrobial resistance (AMR) genes

From the dataset of CRISPR-Cas genomes (Tables S1-3), we conducted an AMR gene analysis to investigate the potential of self-targeting AMR genes with endogenous CRISPR-Cas systems (Figure 2a). The different types of CRISPR-Cas genomes, except for Type IA and Type IV, had several AMR-related genes per genome, ranging from 0.3 genes per genome for Type VA to 23.5 genes per genome for Type IE (Figure 2b). AMR-related genes were absent in Types IA and IV because they only had few CRISPR-Cas genomes that belonged to nonpathogenic prokaryotes, such as *Clostridium perfringens* and *Alteromonas mediterranea*. In the reference gene catalog of the AMR database [26], the AMR-related genes are further classified into antimicrobial resistance, stress response and virulence genes. The classification results show that most genes give antimicrobial resistance, and there are a few genes that confer virulence to the pathogens and others respond to external stresses such as metal or biocide (Table 1). It is intriguing to observe that only certain types of the CRISPR-Cas genomes (Types IB, IE, IF, IIC and IIIA) have virulence genes, with Type IE having the highest ratio of virulence to AMR genes. Many CRISPR-Cas genomes of Type IE belong to pathogenic strains, including *Salmonella enterica* and *Shigella* spp. that are on the WHO priority pathogens list for new antibiotics. This result shows that the co-occurrence of AMR-related genes and CRISPR-Cas systems differ vastly depending on the Cas system type, thus the AMR analysis is the first step to understand the genomic background of target pathogens to achieve effective design and delivery of CRISPR-based antimicrobials.

### CRISPR-Cas genomes have diverse DNA repair pathways

We investigated the distribution of DNA repair pathways in the CRISPR-Cas genomes, based on the previous study of double-strand break (DSB) repair pathways in prokaryotic genomes [27]. We searched diverse DSB repair pathways, including the SOS response, the non-homologous end-joining (NHEJ), and various nuclease proteins. Each DNA repair pathway per genome was calculated for the CRISPR-Cas genomes of each Cas system type (Table S4). The results are visualized as a heatmap (Figure 3) with the proteins belonging to each DNA repair pathway shown on the right axis label (e.g. Ku, LigD1, LigD2 and LigD3 are components of NHEJ pathways). The heatmap shows that some DSB repair pathways are enriched in most CRISPR-Cas genomes, including the AddAB pathway, AdnAB pathway and RuvAB pathway. Furthermore, some proteins such as RecG and RecN are enriched in almost all types of the CRISPR-Cas genomes.

The DSB repair pathways of some Cas system types show outlier patterns to the other CRISPR-Cas genomes (Figure 3). Particularly, the DSB repair pathways of Type ID stand out as an outlier, in which the RecBCD and the RuvAB pathways are more enriched while the AddAB pathway is less enriched, relative to the other types. Additionally, the CRISPR-Cas genomes of Type IIIA and Type IV stand out as outliers to have relatively high numbers of genes belonging to the NHEJ pathway, which have only been recently identified and verified to activate in prokaryotic genomes [28,29]. For this pathway, ligation is usually carried out by LigD proteins, but other ligases can be recruited by Ku in their absence.

## DNA repair during acquisition generates variant CRISPR repeats

Recent studies on the acquisition step shows that the site-integration of new spacers in CRISPR arrays is polarized; most spacers are added to the leader end of the CRISPR array [23–25] (Figure 4a). In this step, the Cas1-Cas2 complex acts as a spacer integrase [30,31], during which the terminal 3' ends of a protospacer catalyzes a nucleophilic attack on each end of the repeat. After this reaction, the 3' ends of the protospacer are ligated to the repeat ends and the single-strand gaps are presumed to be duplicated by a DNA polymerase [32–34]. During this repeat duplication, the repeat sequence at the leader end of the CRISPR array is used as a template due to the polarity of the spacer acquisition.

We investigated the CRISPR repeats of each Cas system type by dimensionality reduction to visualize the variation of CRISPR repeat sequences resulting from the DNA repair activities (Figure 4b). We used various summary statistics of biological features to interpret the principal components of these clusters. Each cluster of the repeats differs in mean length and standard deviation (Table 2 and S5). The cluster analysis shows the length of a sequence and the metric entropy (i.e. randomness of a sequence) are captured on the first latent dimension (Figures S1 and S2). Furthermore, the clusters have a wide range of GC/AT ratio, which is captured on the second latent dimension (0.66 of Cluster 0 vs. 2.19 of Cluster 1). Another important feature of the CRISPR repeats is the RNA secondary structure. The clusters of low minimum free energy (Cluster 1 and 4) lie on the upper side of the second principal component, which indicates highly structured CRISPR repeats. Contrarily, those with the high minimum free energy (Cluster 0 and 2) lie on the lower side of the second principal component, which indicates CRISPR repeats without distinct secondary structure.

## CRISPR repeat structures show the patterns of DNA repair by the Cas system type

CRISPR arrays contain multiple repeats that separate unique spacers (typically, <50 spacers in bacteria and <100 spacers in archaea) [35], and the dimensionality reduction study showed the variation of these repeats within an array. To elucidate how these secondary structures of CRISPR repeats change over time due to DNA repair during the acquisition, we predicted RNA secondary structures of individual repeats within an array and quantified the Minimum Free Energy (MFE) associated with the secondary structure (Tables S6 and S7). The lower the MFE value, the higher the probability of sequences forming stable RNA secondary structures. We plotted the time-series graphs of the MFE values for CRISPR repeats within each array chronologically, in which the CRISPR repeats were separated by the number of unique repeats in an array (Figure S5). The number of unique repeats was assumed to be mutation events during the spacer acquisition process, varying from 2 to 24 time points. These time-series graphs show that the MFE values of CRISPR repeats fluctuate over time. This result shows that the secondary structures of CRISPR repeats are dynamic due to mutation events during the spacer acquisition process. Another noticeable trend is the difference in the baseline of MFE values in CRISPR repeats associated with different Cas system types. For example, the MFE baselines of Class 2 subtypes, including IIA, IIB, and IIC, were consistently higher than some of Class 1 subtypes, including IC, IE, and IF. Interestingly, the MFE baselines of IA, IB and some III types do not appear to follow the same trend.

To visualize the change in the CRISPR repeat structure over time, we built a selected collection of the graphical output of these RNA structures by the associated Cas system type (Figure S6). Consistent with the time-series graphs built with the MFE values, the CRISPR repeat structures of Class 1 subtypes, particularly IC, IE, and IF, tend to have more distinctive hairpin structures of palindromic sequences over time as compared to those of Class 2 subtypes. Such difference in time-series patterns of CRISPR secondary structures according to the associated Cas system types raises an intriguing question of the differential effects of DNA repair during the genome-editing events of CRISPR-Cas systems.

## Simulated studies show the effects of DNA repair under Lamarckian evolution

We simulated a selection of CRISPR repeats associated with Class 1 Type IE and Class 2 Type IIA (Table S6) using the population genetic model that simulates genetic drift of mutations. These simulation studies of the Darwinian evolution model were conducted to compare the evolution of CRISPR repeats that undergo genome-editing events equivalent to Lamarckian evolution [21]. According to the population genetic model, mutations on non-coding sequences are assumed to be neutral and their genetic drift through generations is modeled through binomial sampling [36,37]. As shown in Figure 5a, the CRISPR repeats associated with Class 1 Type IE maintain low MFE values temporally despite some fluctuations. However, the simulated trajectory of MFE values from the input repeats of the same initial sequences shows a trend towards zero MFE (Figure 5b). The difference in these trends is highlighted by the visualization of RNA secondary structures under each graph. Under the population genetic model, any mutation on the CRISPR repeat sequences is likely to degrade the RNA secondary structure by breaking the palindromic patterns. However, the CRISPR repeats associated with Class 1 Type IE tend to maintain the RNA secondary structures in the presence of mutations more robustly than expected. For the CRISPR repeats associated with Class 2 Type IIA (Figure 5c), the temporal patterns in MFE values are similar to the simulated patterns of MFE from the same initial sequences (Figure 5d). These temporal patterns are consistent as the initial repeat sequences of Type IIA are unstructured, thus mutations cannot break down the RNA secondary structure.

## Discussion

CRISPR-Cas systems were initially discovered in prokaryotic genomes, which was found to be an adaptive immunity against invading mobile genetic elements. Due to their ability to cut DNA/RNA specifically with the CRISPR RNA as a guide template, CRISPR-Cas systems were first applied as genome-editing tools to alter certain phenotypic features in eukaryotes, including somatic human cells and agricultural plant cells. Recently, CRISPR-based antimicrobials are being repurposed as a highly potent alternative to traditional antibiotics to self-target drug-resistant pathogens [8,38,39]. The CRISPR RNA component can be reprogrammed to self-target antimicrobial resistance (AMR) genes in the chromosome or plasmid of these drug-resistant pathogens. Moreover, CRISPR-based antimicrobials have the potential to be used as preventive measures, such as controlling reservoirs of AMR genes in microbial communities to regain or retain the antimicrobial activity of traditional antibiotics [13]. However, most prokaryotic genomes have the ability to repair DNA damage, which includes the nuclease activity of CRISPR-Cas systems that requires DNA repair to integrate new spacers and to regenerate new repeats in CRISPR arrays [27,40].

According to the comprehensive survey of AMR-related genes in the curated prokaryotic genome dataset, most CRISPR-Cas genomes (except for Types IA and IV) have numerous AMR-related genes that can be self-targeted with endogenous CRISPR-Cas systems. This co-occurrence of CRISPR-Cas systems and AMR-related genes enable the delivery of CRISPR-based antimicrobials to be simplified to self-targeting CRISPR arrays on mobile genetic elements. Recently, phage capsids have been engineered to deliver self-targeting CRISPR-based antimicrobials to pathogenic bacteria [14,15]. For pathogens with both CRISPR-Cas systems and AMR-related genes, a simpler construct of self-targeting CRISPR arrays can be packaged into these viral vectors [16]. Efficient delivery to specific bacteria is one of the main challenges of programmable CRISPR-based antimicrobials. Although several studies demonstrated genetic elements encoding foreign systems can be delivered to target bacteria using several vectors such as phage capsids, conjugative plasmids and nanoparticles [10], the specificity and efficiency of such delivery vectors in a complex natural environment is still an ongoing area of research. Furthermore, the defense mechanisms and the resistance development of pathogens against these CRISPR-based antimicrobials should be studied and monitored extensively to demonstrate the long-term effectiveness of these novel antibiotics [8,38,39].

In this study, we investigated the potential interference of DNA repair pathways in utilizing CRISPR-based antimicrobials. Given that we found numerous and diverse DNA repair pathways in the CRISPR-Cas genomes, we focused on two general mechanisms to repair DNA damage.

Homologous recombination (HR) requires a homologous template to repair the DNA damage with high-fidelity [18,40]. We found that all CRISPR-Cas genomes have diverse HR-related genes, including genes necessary for RecBCD, AddAB and AdnAB pathways. Many bacteria contain multiple copies of the genome, or at least partially replicated forms before cell division, which may require CRISPR-based antimicrobials to perform simultaneous targeting due to the presence of diverse HR pathways. Non-homologous end-joining (NHEJ) is a DNA repair pathway that processes the DNA damage and directly ligates the DNA ends without requiring template DNA [18]. Previously, bacteria were assumed to rely mainly on homologous recombination (HR) to repair double-strand breaks, but recent discovery of alternative non-homologous end-joining pathways strengthens the evidence that bacteria have the ability to ligate unrelated DNA ends that do not share homology to create new genetic combinations [28]. However, Type IIA CRISPR-Cas systems in bacteria were found to inhibit NHEJ repair pathways due to the antagonistic interactions of recognizing the same DNA damage [40]. Consistently, we found that CRISPR-Cas genomes of Type IIA are void of NHEJ-related genes. However, we found that other CRISPR-Cas genomes have NHEJ-related genes, with Type IIIA having been relatively enriched. These findings show the complex interactions between CRISPR-Cas systems and DNA repair pathways in CRISPR-Cas genomes, and the application of CRISPR-based antimicrobials on bacteria require extensive investigations on the genomic background of target bacteria.

Inspired by the interplay between CRISPR-Cas systems and DNA repair pathways, we further investigated the unique genome-editing features governing the evolution of CRISPR-Cas genomes. The ability of CRISPR-Cas immunity to specifically modify the genome of a prokaryote in response to an external challenge (e.g. virus infection) has been recognized as an unique example of Lamarckian evolution [21]. Unlike Darwinian evolution whose variation results from random mutations, Lamarckian evolution relies on the high specificity of mutations that results in an efficient adaptation to the external challenge, and the necessity to co-evolve effective DNA repair pathways along with CRISPR-Cas systems was predicted by theoretical evolutionary modeling [41]. In this study, we brought further insights into the interaction between CRISPR-Cas systems and DNA repair pathways by time-series visualization of CRISPR repeat secondary structures and the simulation studies of CRISPR repeat evolution. We demonstrated that the diversity of CRISPR repeat structures is an important biological feature of different CRISPR-Cas systems, and the variation within a CRISPR array reflects the interplay of CRISPR-Cas systems and DNA repair pathways during the genome-editing event of spacer acquisition. Furthermore, the simulation studies elucidated that the secondary RNA structures of Type I CRISPR repeats are maintained better than expected under Darwinian evolution, which further elucidates the ability of some CRISPR-Cas genomes to repair DNA damage with high fidelity.

From this study, we emphasized the importance of understanding the genomic background of CRISPR-Cas genomes to exploit the potential of CRISPR-based antimicrobials to self-target AMR-related genes. CRISPR-based antimicrobials are unique programmable tools that can target bacteria specifically for their pathogenicity, despite the various challenges such as delivery issues and host resistance. We are currently in urgent need of next-generation antibiotics. The antibiotic market is currently not viable as new antibiotics can only be used sparingly as the last resort to prevent the rise of new drug resistance [42–44]. As opposed to the traditional antibiotics, for which drug resistance emerges rapidly, CRISPR-based antimicrobials offer an opportunity to exploit the recent progress in understanding the complexity and evolution of prokaryotic genomes to strategically counteract against the spread of drug-resistant bacteria.

## Methods

### Curating a labeled dataset of CRISPR-Cas genomes by the Cas system type

We used a public database CRISPRCasdb (downloaded on 21/01/2021) to build a dataset of CRISPR-Cas genomes labeled by the Cas system type, which we define as prokaryotic genomes that have one complete set of Cas genes and one associated CRISPR array. We chose this one-to-one association to eliminate potential inaccuracy resulting from mislabeling associations between multiple CRISPR arrays and multiple Cas gene systems within the same genome. From 26,340 bacterial genomes and 436 archaeal genomes, CRISPRCasdb found 10,890 (41.34%) bacterial genomes with CRISPR arrays and 333 (76%) archaeal genomes with CRISPR arrays (Table S1). Overall, 9,554 (36.27%) bacterial genomes and 308 (70.74%) archeal genomes had both CRISPR arrays and Cas gene systems. We, hereinafter, refer to CRISPR arrays in prokaryotic genomes without Cas gene systems as "orphan arrays". As each CRISPR array typically contains multiple repeat sequences, the total number of unique repeats adds up to 26,958.

The number of non-redundant CRISPR-Cas genomes labeled by associated Cas system type from the CRISPRCasdb is summarized in Table S2. The number of CRISPR-Cas genomes varies by the Cas system type. For example, there are 209 CRISPR-Cas genomes associated with Type IE, whereas only 1 CRISPR-Cas genome is associated with Type VIB2. The disparity in the types may be due to CRISPRCasdb having biased sampling for human pathogens. Furthermore, this may result from other factors such as the selection criterion of those with one-to-one associations, the recent discovery of some subtypes (such as Type VI), and potentially their relative rarity in nature. The number of unique CRISPR repeats labeled by different Cas system types is shown in Table S3. For visualization analyses, we merged the CRISPR-Cas genomes associated with the Cas system types that are extremely rare into one category (labeled as "ex"), while keeping the subtypes of Type I, II, and III as separate categories.

### Analysis of AMR genes and DNA repair pathways in CRISPR-Cas genomes

For the AMR gene analysis, we used the NCBI Antimicrobial Resistance Gene Finder [26] that has an accompanying database of antimicrobial resistance genes, including some point mutations (AMRFinderPlus Version 3.10.20). We ran this software with protein sequences of the CRISPR-Cas genomes to search for AMR-related genes, which uses BLASTP and HMMER for gene matches and classification of novel sequences by building a hierarchical tree of gene families.

For the DNA repair analysis, we used the components of the double-strand break repair system that had previously been constructed using MacSyFinder (Version 1.0.2) [27]. From these DNA repair pathways, the protein profile for new proteins had been built with the multiple alignment sequence of homologous proteins using MAFFT (Version 7.205) and HMMER (Version 3.1) [27]. We downloaded the whole genomes which contained each CRISPR array by the associated Cas system from NCBI (downloaded 10/01/2022), and we used the HMM profiles of the DSB repair system to search for the components with HMMsearch (Version 3.3.2). We counted the number of each component in the DSB repair system above the sequence reporting threshold (E-value $> 1e^{-3}$) and calculated the number of each component per genome for each CRISPR array by the associated Cas system.

### Dimensionality reduction of CRISPR repeats

Principal Component Analysis (PCA) reduces the dimensions of data by computing the principal components and uses the first few to increase the interpretability. We used a PCA approach that transfers the sequence matrix to a boolean vector for direct analysis of nucleotide sequences [45]. Featurization of nucleotide sequences has been explored extensively in previous studies, mainly through encoding the four nucleotides with one-hot vectors [46–49]. This transformation of nucleotide sequences has merits that it is completely reversible, and PCA can be directly applied to the transformed sequence matrix. The maximum length of repeats for all the categories is 50 (Figure S1). For interpretability, we used a 2-dimensional latent space, as the third dimension does not add additional information about the biological features for this study (data not shown).

### Clustering with Gaussian Mixture Models (GMM)

We used Gaussian Mixture Models (GMM) as a probabilistic model to define clusters. GMMs assume all data points follow a mixture of Gaussian distributions, with a fixed number of unknown parameters. GMMs are a generalized k-means clustering that incorporates the centers of Gaussian distributions and the covariance structure of input data. GMMs need the number of clusters to be pre-defined before using the algorithm. For model selection, we used the Bayesian information criterion (BIC) to choose the number of clusters without overfitting [50]. The BIC introduces a penalty term for the increasing number of parameters in the model:

$$BIC = k * ln(n) - 2 * ln(\hat{L})$$

where $k$ is the number of parameters, $n$ is the observed data, and $\hat{L}$ is the maximized value of the likelihood function of the evaluated model.

Using Gaussian Mixture Models (GMM) as a probabilistic model, we evaluated a range of cluster numbers (1 to 9), with four different covariances of input data for each model (spherical, tied, diagonal, and full). The BIC scores from the GMM model selection for the repeats is summarized in Figure S3. The BIC scores reveal that assuming the full covariance of input data renders the best result in every model. For the GMM models with the full covariance, the last BIC score to drop significantly occurs between the clusters of 4 and 5. Thus, the GMM model with 5 clusters was chosen as the simplest GMM model that best fits this data according to the BIC criterion (Figure S4). According to the GMM model, we designated each cluster with the associated Cas system type for further analyses (Table S5).

### Biological feature interpretations of clusters

We evaluated each cluster with summary statistics to infer biological interpretations of the features the PCA extracted from the CRISPR repeats (Table 2). We calculated the entropy of the CRISPR repeats from each cluster to assess the randomness in these sequences. We used the Shannon entropy bounded between 0 and 1 as a measure of information content in a sequence [51]:

$$H(X) = -\sum_{i}^{M} P(x_i) \, log_2 \, P(x_i)$$

where $P(x_i)$ is the probability of the event $x_i$. The Shannon entropy gives the maximum entropy for equiprobable and independent states of the four nucleotides (A, T, G, C). We obtained the metric entropy by dividing the Shannon entropy by the sequence length (Table 2).

We used the ViennaRNA Package to predict the RNA secondary structure of the CRISPR repeats. The RNAfold (Version 2.4.14) function of the package calculates the minimum free energy (MFE in kcal/mol) of the thermodynamic ensemble to predict the stability of RNA secondary structures [52]. We chose the centroid method to predict the optimal secondary structure, which results in the secondary structure with a minimum total base-pair distance to the entire thermodynamic ensemble of structures [52,53]. The centroid method finds the optimal secondary structure that minimizes the following sum of minimum base-pair distances:

$$\sum_{l \le k \le m} \sum_{i} \sum_{j} (I_{ij}^k - I_{ij})^2$$

for a set of $m$ secondary structures $I_1, I_2, \ldots, I_m$, with $I_k = \{I_{ij}^k\}$, $1 \leq k \leq m$. The biological features of CRISPR repeats, including metric entropy, sequence length, GC/AT ratio, and minimum free energy, were calculated by the clusters modeled by GMM.

### Time-series patterns in RNA secondary structures of CRISPR repeats

To visualize the secondary structures of CRISPR repeats, the Vienna RNA software (Version 2.4.18) was used. Using the software, minimum free energy (MFE) values for RNA secondary structures were predicted [54], where an optimal secondary structure among the centroid structure, the partition function, and the matrix of base pairing probabilities [55] was recorded. The MFE values of the optimal secondary structure were obtained for all CRISPR repeats, and they were plotted in time-series graphs by the number of time points in each CRISPR array (Figure S5). For the visualization of RNA secondary structures, 100 CRISPR repeats were selected randomly to ensure every species of bacteria was included for the subtypes with many sequences (>100). Otherwise, all repeats in the dataset were analyzed for the subtypes with 100 or less sequences (Figure S6).

### Simulated patterns of the Minimum Free Energy (MFE) of CRISPR repeats

To investigate the time-series patterns of CRISPR secondary structures under Lamarckian evolution, we simulated the evolution of CRISPR repeats under Darwinian evolution of genetic drift. We chose CRISPR repeats of the two subtypes (Class 1 Type IE and Class 2 Type IIA) that had the most prominent patterns from our previous time-series analyses for simulation studies. We chose CRISPR repeats that had 5 time points in the arrays to show clear temporal trends and only those arrays with the known direction (Table S6). The CRISPR repeats sequences from the first time point were the input sequences to the following simulation studies. For the simulation, we assumed the following population genetic model: genetic drift of mutations under binomial sampling of wildtype and mutant between generations. The mutation rate of microbes in nature is extremely difficult to measure, thus we chose the high end of the estimated range of mutation rates in microbial organisms ($1e^{-5}$ mutation per generation). For every mutation event, one of the four nucleotides (A, U, G, C) was randomly chosen to replace the wildtype nucleotide. To ensure the presence of mutations, we ran the simulation for 10,000 generations, and these simulations were run for 5 time points. The simulated output of CRISPR repeat sequences of each time point was processed using Vienna RNA software (Version 2.4.18) as above for visualization of RNA secondary structures and quantification of MFE values. We repeated these simulations 100 times for each input sequence of CRISPR repeats, and the means of MFE values were plotted in time-series graphs for comparison (Figure 5).

# Tables

**Table 1.** Classification of the AMR-related genes in the CRISPR-Cas genomes by the Cas system type.

| Cas system type | AMR | Virulence | Stress: Acid | Stress: Biocide | Stress: Metal | Stress: Heat |
|---|---|---|---|---|---|---|
| **IB** | 137 | 8 | - | - | 36 | - |
| **IC** | 546 | - | - | 3 | 115 | 5 |
| **ID** | - | - | - | - | 2 | - |
| **IE** | 1631 | 1351 | 438 | 92 | 1356 | 45 |
| **IF** | 571 | 42 | 24 | 19 | 306 | 1 |
| **IIA** | 313 | - | - | - | 89 | - |
| **IIB** | 26 | - | - | - | - | - |
| **IIC** | 918 | 40 | - | 37 | 194 | - |
| **IIIA** | 416 | 60 | 1 | 2 | 32 | - |
| **IIIB** | 32 | - | - | 1 | 5 | - |
| **IIIC** | 1 | - | - | - | - | - |
| **IIID** | 5 | - | - | - | 3 | - |
| **VA** | 3 | - | - | - | - | - |
| **VIB1** | 2 | - | - | - | 1 | - |
| **VIB2** | 2 | - | - | - | - | - |

**\*** CRISPR-Cas genomes of Type IA and Type IV had no AMR-related gene.

**Table 2.** Summary statistics of CRISPR repeats by the Gaussian Mixture Model cluster.

| Cluster | Mean length ± s.d. (number of data) | GC/AT ratio ± s.d. | Metric entropy ± s.d. | Minimum free energy of RNA ± s.d. |
|---|---|---|---|---|
| **0** | 37.35±3.38 (n = 2,753) | 0.66±0.42 | 0.050±0.0047 | -4.13±4.57 |
| **1** | 28.97±0.22 (n = 2,122) | 2.19±0.52 | 0.065±0.0021 | -13.57±2.10 |
| **2** | 30.44±2.11 (n = 1,449) | 0.93±0.67 | 0.061±0.0053 | -6.73±5.77 |
| **3** | 27.59±1.64 (n = 3,007) | 1.46±0.75 | 0.070±0.0050 | -8.86±4.17 |
| **4** | 32.72±2.61 (n = 1,759) | 1.69±0.60 | 0.058±0.0059 | -11.53±3.39 |

**Figure 1.** The genomic background analysis of CRISPR-Cas genomes.

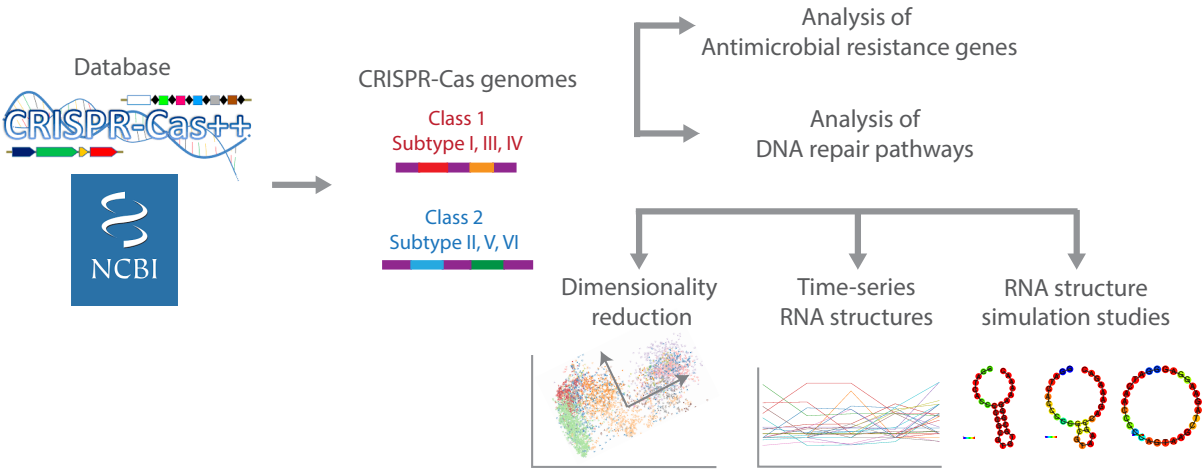

**Figure 2.** (a) The first bar plot summarizes the CRISPR-Cas genomes in the dataset by each Cas system type, with the brown color representing bacterial genomes and green color representing archaeal genomes. The 3D macromolecular protein structure of a signature Cas protein for each system is shown on the left panel. (b) The second bar plot shows the number of AMR-related genes per CRISPR-Cas genome in the dataset by each Cas system type.

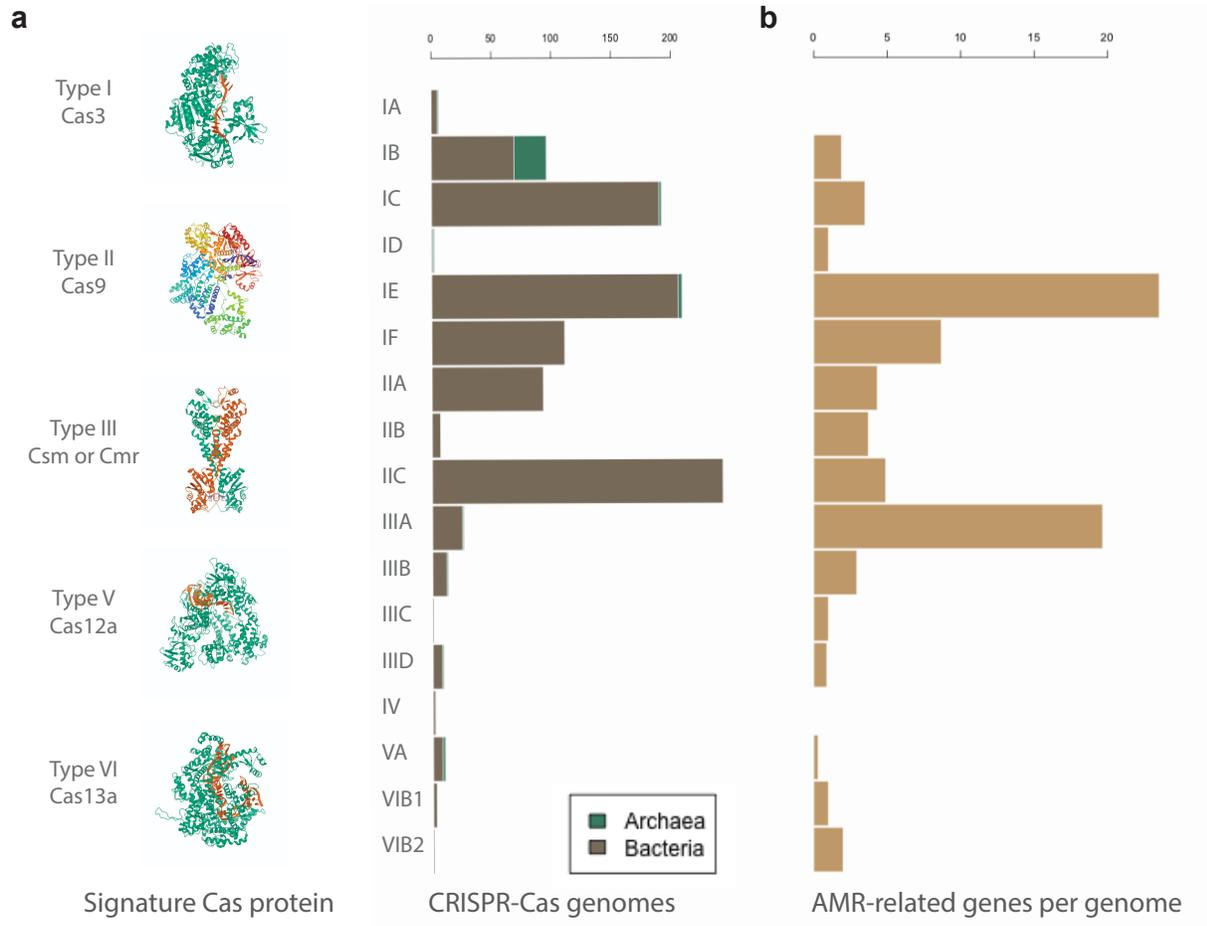

**Figure 3.** Heatmap showing the number of DNA repair pathways per CRISPR-Cas genome for each Cas system type. The name of the protein belonging to each DNA repair pathway is indicated on the right axis label. The color bar shows a scale from 0 - 1.7 DNA repair proteins per CRISPR-Cas genome, with the red color indicating the highest frequency.

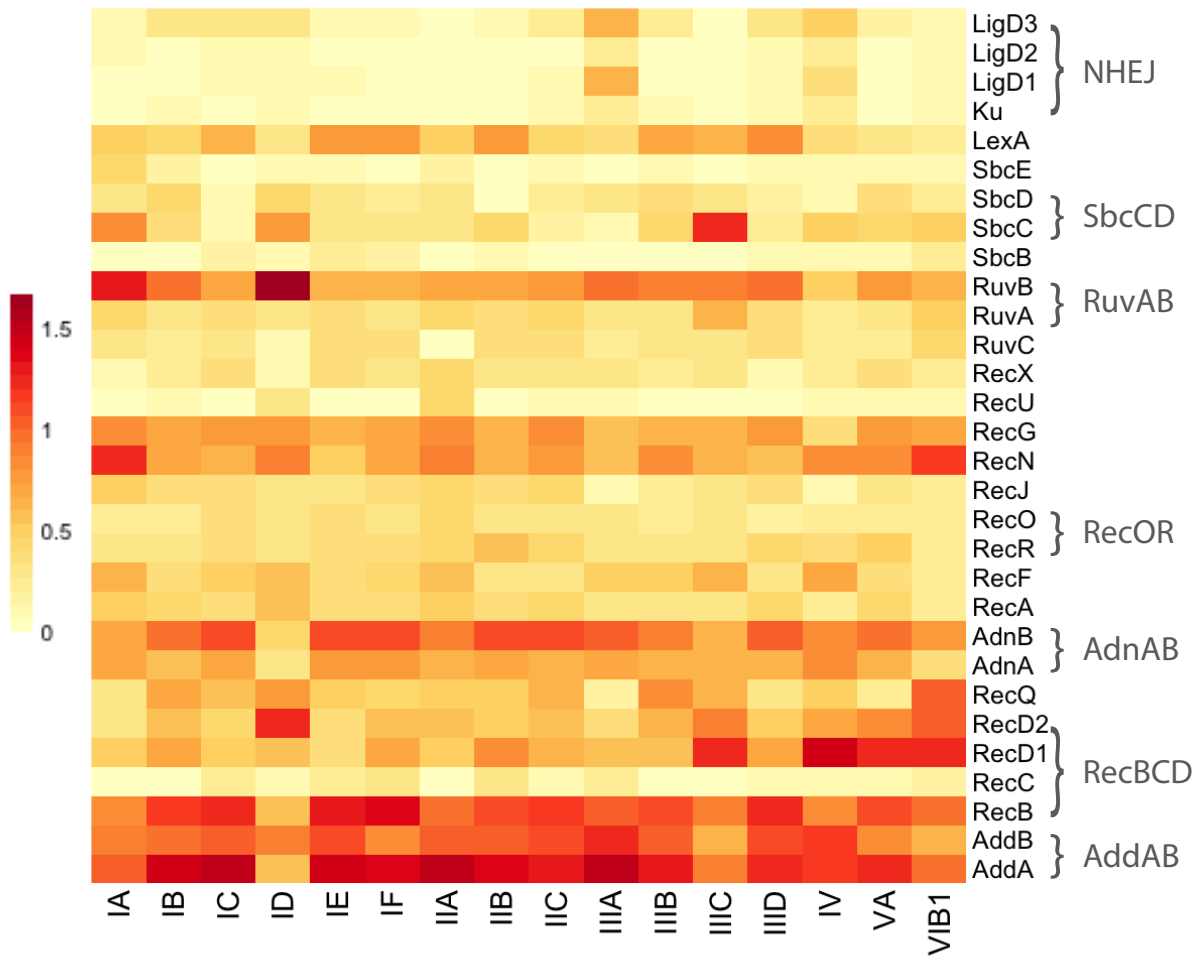

**Figure 4.** (a) Acquisition steps of new spacers in a CRISPR array show how repeats are being repaired by the DNA repair pathways after new spacer acquisition. (b) Projection of CRISPR repeats on the 2-dimensional latent space labeled with the associated Cas system type.

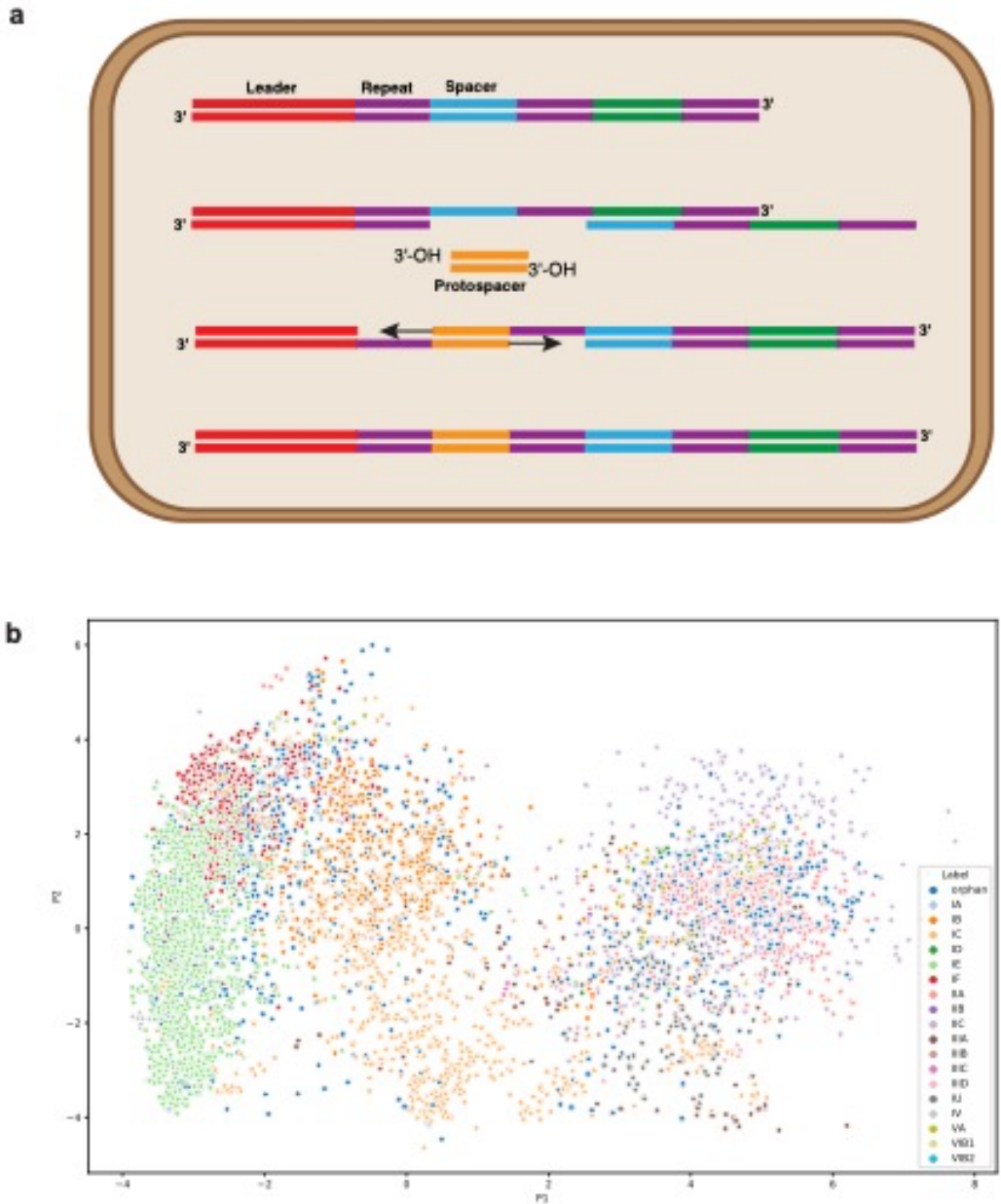

**Figure 5.** Time-series graphs of the secondary structure of CRISPR repeats in the forward direction with 5 time points. (a) Minimum free energy of Class 1 Type IE CRISPR repeats. (b) Simulated minimum free energy of Class 1 Type IE CRISPR repeats. (c) Minimum free energy of Class 2 Type IIA CRISPR repeats. (d) Simulated minimum free energy of Class 2 Type IIA CRISPR repeats.

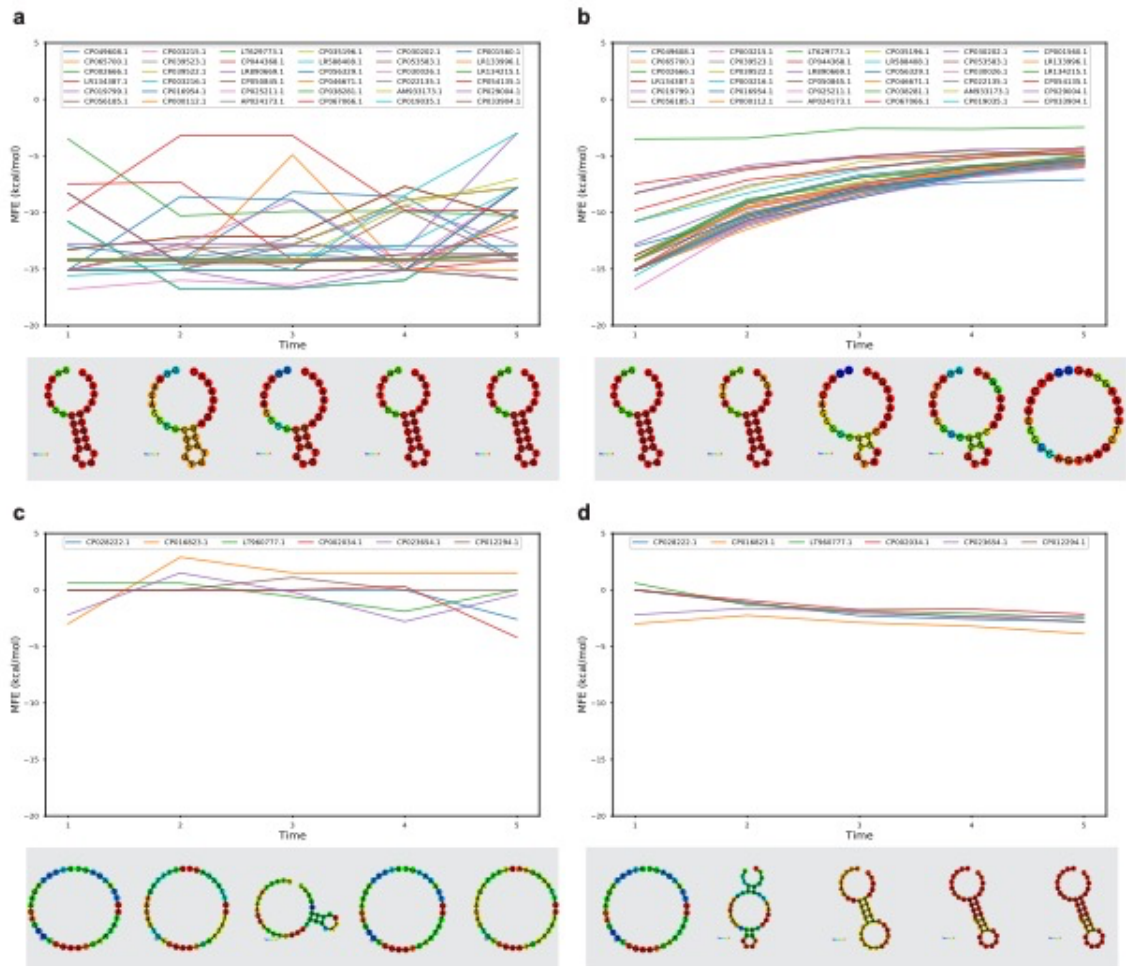

## Data and code availability

All the CRISPR-Cas sequences are available in the CRISPR-Cas++ database (https://crisprcas.i2bc.paris-saclay.fr) as well as our project GitHub page (https://github.com/hjshim/CRISPR_DR). For AMR analysis, we used AMRFinderPlus v.3.10.20 (https://github.com/ncbi/amr). For DNA repair analysis, we used HMM search v.3.3.2 (http://hmmer.org/). For dimensionality reduction, we used Direct-PCA (https://github.com/TomokazuKonishi/direct-PCA-for-sequences) and scikit-learn (https://scikit-learn.org). For RNA secondary structure, we used Vienna RA software v.2.4.18. For all data analysis and visualization, Python v.3.7.3 (https://www.python.org), SciPy v.1.1.0 (https://www.scipy.org), seaborn v.0.9.0 (https://github.com/mwaskom/seaborn) were used.

## Competing interests

The authors declare no competing interests.

## Supplementary Information

Supplementary Information will be provided upon request.

Rev Microbiol. 2013;11: 9–13.